\pdfoutput=1
\documentclass[12pt,a4paper]{article}
\usepackage{ifthen} 
\newboolean{pdflatex}
\setboolean{pdflatex}{true} 

\newboolean{articletitles}
\setboolean{articletitles}{true} 

\newboolean{uprightparticles}
\setboolean{uprightparticles}{false} 

\def\paperauthors{LHCb collaboration} 
\def\paperasciititle{A search  for rare B->Dmumu decays}
\def\papertitle{A search  for rare \GenDecay decays} 
\def\paperkeywords{{High Energy Physics}, {LHCb}} 
\def\papercopyright{\the\year\ CERN for the benefit of the LHCb collaboration} 
\def\paperlicence{CC BY 4.0 licence}
\def\paperlicenceurl{https://creativecommons.org/licenses/by/4.0/}

\usepackage[top=1in, bottom=1.25in, left=1in, right=1in]{geometry}

\usepackage{microtype}
\usepackage{lineno}  
\usepackage{xspace} 
\usepackage{caption} 

\usepackage{graphicx}  
\usepackage{color}
\usepackage{colortbl}
\graphicspath{{./figs/}} 

\usepackage{amsmath} 
\usepackage{amssymb}
\usepackage{amsfonts}
\usepackage{upgreek} 

\newcommand*\patchAmsMathEnvironmentForLineno[1]{%
\expandafter\let\csname old#1\expandafter\endcsname\csname #1\endcsname
\expandafter\let\csname oldend#1\expandafter\endcsname\csname
end#1\endcsname
 \renewenvironment{#1}%
   {\linenomath\csname old#1\endcsname}%
   {\csname oldend#1\endcsname\endlinenomath}%
}
\newcommand*\patchBothAmsMathEnvironmentsForLineno[1]{%
  \patchAmsMathEnvironmentForLineno{#1}%
  \patchAmsMathEnvironmentForLineno{#1*}%
}
\AtBeginDocument{%
\patchBothAmsMathEnvironmentsForLineno{equation}%
\patchBothAmsMathEnvironmentsForLineno{align}%
\patchBothAmsMathEnvironmentsForLineno{flalign}%
\patchBothAmsMathEnvironmentsForLineno{alignat}%
\patchBothAmsMathEnvironmentsForLineno{gather}%
\patchBothAmsMathEnvironmentsForLineno{multline}%
\patchBothAmsMathEnvironmentsForLineno{eqnarray}%
}

\usepackage{hyperxmp}

\usepackage[pdftex,
            pdfauthor={\paperauthors},
            pdftitle={\paperasciititle},
            pdfkeywords={\paperkeywords},
            pdfcopyright={Copyright (C) \papercopyright},
            pdflicenseurl={\paperlicenceurl}]{hyperref}

\usepackage[colorinlistoftodos,textsize=scriptsize]{todonotes}

\usepackage[bottom,flushmargin,hang,multiple]{footmisc}

\usepackage[all]{hypcap} 

\usepackage{xspace} 
\usepackage{upgreek}

\def\lhcb   {\mbox{LHCb}\xspace}

\def\babar  {\mbox{BaBar}\xspace}

\def\MagUp {\mbox{\em Mag\kern -0.05em Up}\xspace}

\ifthenelse{\boolean{uprightparticles}}%
{
 
 \def\Pgamma      {\ensuremath{\upgamma}\xspace}

 \def\Pmu         {\ensuremath{\upmu}\xspace}                 
 \def\Pnu         {\ensuremath{\upnu}\xspace}                 
                  
 \def\Ppi         {\ensuremath{\uppi}\xspace}

 \def\Ppsi        {\ensuremath{\uppsi}\xspace}

 \def\PDelta      {\ensuremath{\Delta}\xspace}                 
 \def\PXi         {\ensuremath{\Xi}\xspace}                 
 \def\PLambda     {\ensuremath{\Lambda}\xspace}                 
 \def\PSigma      {\ensuremath{\Sigma}\xspace}                 
 \def\POmega      {\ensuremath{\Omega}\xspace}                 
 \def\PUpsilon    {\ensuremath{\Upsilon}\xspace}
 \let\oldPi\Pi
 \def\PPi         {\ensuremath{\oldPi}\xspace}

 \def\PB      {\ensuremath{\mathrm{B}}\xspace}                 
                  
 \def\PD      {\ensuremath{\mathrm{D}}\xspace}

 \def\PJ      {\ensuremath{\mathrm{J}}\xspace}                 
 \def\PK      {\ensuremath{\mathrm{K}}\xspace}

 \def\PW      {\ensuremath{\mathrm{W}}\xspace}

 \def\Pb      {\ensuremath{\mathrm{b}}\xspace}                 
 \def\Pc      {\ensuremath{\mathrm{c}}\xspace}

 \def\Pi      {\ensuremath{\mathrm{i}}\xspace}

 \def\Ps      {\ensuremath{\mathrm{s}}\xspace}

 \def\thebaroffset{0.0em}
}
{
 
 \def\Pgamma      {\ensuremath{\gamma}\xspace}

 \def\Pmu         {\ensuremath{\mu}\xspace}                 
 \def\Pnu         {\ensuremath{\nu}\xspace}                 
                  
 \def\Ppi         {\ensuremath{\pi}\xspace}

 \def\Ppsi        {\ensuremath{\psi}\xspace}                 
                  
 \mathchardef\PDelta="7101
 \mathchardef\PXi="7104
 \mathchardef\PLambda="7103
 \mathchardef\PSigma="7106
 \mathchardef\POmega="710A
 \mathchardef\PUpsilon="7107
 \mathchardef\PPi="7105
                  
 \def\PB      {\ensuremath{B}\xspace}                 
                  
 \def\PD      {\ensuremath{D}\xspace}

 \def\PJ      {\ensuremath{J}\xspace}                 
 \def\PK      {\ensuremath{K}\xspace}

 \def\PW      {\ensuremath{W}\xspace}

 \def\Pb      {\ensuremath{b}\xspace}                 
 \def\Pc      {\ensuremath{c}\xspace}

 \def\Pi      {\ensuremath{i}\xspace}

 \def\Ps      {\ensuremath{s}\xspace}

 \def\thebaroffset{0.18em}
}
\newcommand{\offsetoverline}[2][\thebaroffset]{\kern #1\overline{\kern -#1 #2}}%

\makeatletter
\ifcase \@ptsize \relax
  \newcommand{\miniscule}{\@setfontsize\miniscule{4}{5}}
\or
  \newcommand{\miniscule}{\@setfontsize\miniscule{5}{6}}
\or
  \newcommand{\miniscule}{\@setfontsize\miniscule{5}{6}}
\fi
\makeatother

\DeclareRobustCommand{\optbar}[1]{\shortstack{{\miniscule (\rule[.5ex]{1.25em}{.18mm})}
  \\ [-.7ex] $#1$}}

\def\mup        {{\ensuremath{\Pmu^+}}\xspace}
\def\mun        {{\ensuremath{\Pmu^-}}\xspace} 

\def\mumu       {{\ensuremath{\Pmu^+\Pmu^-}}\xspace}

\def\ellp       {{\ensuremath{\ell^+}}\xspace}

\def\ellell     {\ensuremath{\ell^+ \ell^-}\xspace}

\def\neub       {{\ensuremath{\overline{\Pnu}}}\xspace}

\def\neumb      {{\ensuremath{\neub_\mu}}\xspace}

\def\g      {{\ensuremath{\Pgamma}}\xspace}

\def\W      {{\ensuremath{\PW}}\xspace}

\def\squark    {{\ensuremath{\Ps}}\xspace}

\def\cquark    {{\ensuremath{\Pc}}\xspace}

\def\bquark    {{\ensuremath{\Pb}}\xspace}

\def\pion   {{\ensuremath{\Ppi}}\xspace}
\def\piz    {{\ensuremath{\pion^0}}\xspace}
\def\pip    {{\ensuremath{\pion^+}}\xspace}
\def\pim    {{\ensuremath{\pion^-}}\xspace}

\def\kaon    {{\ensuremath{\PK}}\xspace}

\def\KorKbar {\kern \thebaroffset\optbar{\kern -\thebaroffset \PK}{}\xspace}

\def\Kp      {{\ensuremath{\kaon^+}}\xspace}
\def\Km      {{\ensuremath{\kaon^-}}\xspace}

\def\Kstarz  {{\ensuremath{\kaon^{*0}}}\xspace}

\def\Dbar    {{\ensuremath{\offsetoverline{\PD}}}\xspace}
\def\D       {{\ensuremath{\PD}}\xspace}

\def\DorDbar {\kern \thebaroffset\optbar{\kern -\thebaroffset \PD}\xspace}
\def\Dz      {{\ensuremath{\D^0}}\xspace}
\def\Dzb     {{\ensuremath{\Dbar{}^0}}\xspace}
\def\Dp      {{\ensuremath{\D^+}}\xspace}
\def\Dm      {{\ensuremath{\D^-}}\xspace}

\def\DpDm    {\ensuremath{\Dp {\kern -0.16em \Dm}}\xspace}

\def\Dstarz  {{\ensuremath{\D^{*0}}}\xspace}
\def\Dstarzb {{\ensuremath{\Dbar{}^{*0}}}\xspace}

\def\Ds      {{\ensuremath{\D^+_\squark}}\xspace}
\def\Dsp     {{\ensuremath{\D^+_\squark}}\xspace}

\def\Dssp    {{\ensuremath{\D^{*+}_\squark}}\xspace}

\def\B       {{\ensuremath{\PB}}\xspace}

\def\BorBbar {\kern \thebaroffset\optbar{\kern -\thebaroffset \PB}\xspace}
\def\Bz      {{\ensuremath{\B^0}}\xspace}

\def\Bd      {{\ensuremath{\B^0}}\xspace}

\def\BdorBdbar {\kern \thebaroffset\optbar{\kern -\thebaroffset \Bd}\xspace}
\def\Bu      {{\ensuremath{\B^+}}\xspace}

\def\Bp      {{\ensuremath{\Bu}}\xspace}

\def\Bs      {{\ensuremath{\B^0_\squark}}\xspace}

\def\BsorBsbar {\kern \thebaroffset\optbar{\kern -\thebaroffset \Bs}\xspace}
\def\Bc      {{\ensuremath{\B_\cquark^+}}\xspace}

\def\Bds     {{\ensuremath{\B_{(\squark)}^0}}\xspace}

\def\jpsi     {{\ensuremath{{\PJ\mskip -3mu/\mskip -2mu\Ppsi}}}\xspace}

\def\Y#1S{\ensuremath{\PUpsilon{(#1S)}}\xspace}

\def\LorLbar     {\kern \thebaroffset\optbar{\kern -\thebaroffset \PLambda}\xspace}

\newcommand{\decay}[2]{\ensuremath{#1\!\to #2}\xspace} 

\def\to                 {\ensuremath{\rightarrow}\xspace}

\def\qsq       {{\ensuremath{q^2}}\xspace}

\def\AT#1     {\ensuremath{A_{\mathrm{T}}^{#1}}\xspace}           

\def\C#1      {\ensuremath{\mathcal{C}_{#1}}\xspace}                       
\def\Cp#1     {\ensuremath{\mathcal{C}_{#1}^{'}}\xspace}                    
\def\Ceff#1   {\ensuremath{\mathcal{C}_{#1}^{\mathrm{(eff)}}}\xspace}        
\def\Cpeff#1  {\ensuremath{\mathcal{C}_{#1}^{'\mathrm{(eff)}}}\xspace}       
\def\Ope#1    {\ensuremath{\mathcal{O}_{#1}}\xspace}                       
\def\Opep#1   {\ensuremath{\mathcal{O}_{#1}^{'}}\xspace}                    

\newcommand{\nospaceunit}[1]{\ensuremath{\text{#1}}}       
\newcommand{\aunit}[1]{\ensuremath{\text{\,#1}}}       

\newcommand{\tev}{\aunit{Te\kern -0.1em V}\xspace}
\newcommand{\gev}{\aunit{Ge\kern -0.1em V}\xspace}
\newcommand{\mev}{\aunit{Me\kern -0.1em V}\xspace}
\newcommand{\kev}{\aunit{ke\kern -0.1em V}\xspace}
\newcommand{\ev}{\aunit{e\kern -0.1em V}\xspace}
 
\newcommand{\mevc}{\ensuremath{\aunit{Me\kern -0.1em V\!/}c}\xspace}
\newcommand{\gevc}{\ensuremath{\aunit{Ge\kern -0.1em V\!/}c}\xspace}
\newcommand{\mevcc}{\ensuremath{\aunit{Me\kern -0.1em V\!/}c^2}\xspace}
\newcommand{\gevcc}{\ensuremath{\aunit{Ge\kern -0.1em V\!/}c^2}\xspace}
\newcommand{\gevgevcccc}{\ensuremath{\gev^2\!/c^4}\xspace} 

\def\mum  {\ensuremath{\,\upmu\nospaceunit{m}}\xspace}

\def\fb   {\ensuremath{\aunit{fb}}\xspace}
\def\invfb   {\ensuremath{\fb^{-1}}\xspace}

\def\ps   {\ensuremath{\aunit{ps}}\xspace}

\def\gsim{{~\raise.15em\hbox{$>$}\kern-.85em
          \lower.35em\hbox{$\sim$}~}\xspace}
\def\lsim{{~\raise.15em\hbox{$<$}\kern-.85em
          \lower.35em\hbox{$\sim$}~}\xspace}

\def\pt         {\ensuremath{p_{\mathrm{T}}}\xspace}

\def\ptot       {\ensuremath{p}\xspace}

\def\evtgen     {\mbox{\textsc{EvtGen}}\xspace}

\def\geant      {\mbox{\textsc{Geant4}}\xspace}

\def\photos     {\mbox{\textsc{Photos}}\xspace}

\def\pythia     {\mbox{\textsc{Pythia}}\xspace}

\def\tell1  {TELL1\xspace}
\def\ukl1   {UKL1\xspace}

\newcommand{\eg}{\mbox{\itshape e.g.}\xspace}
\newcommand{\ie}{\mbox{\itshape i.e.}\xspace}

\newcommand{\lhcborcid}[1]{\href{https://orcid.org/#1}{\hspace*{0.1em}\raisebox{-0.45ex}{\includegraphics[width=1em]{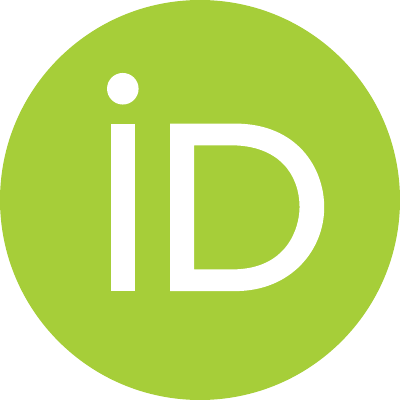}}}}

\usepackage{cite} 
\usepackage{mciteplus}

\usepackage{booktabs}
\usepackage{longtable}
\usepackage{multirow}
\usepackage[section]{placeins}

\newcommand{\fig}[1]{Fig.~\ref{#1}}

\newcommand{\tab}[1]{Table~\ref{#1}}

\newcommand{\BraRat}[1]{\ensuremath{\mathcal{B}\left( #1 \right)}}
\newcommand{\eqinline}[1]{Eq.~\ref{#1}}

\renewcommand{\log}{\log_{10}}
\newcommand{\cls}{\ensuremath{\rm{CL}_{\rm{s}}}\xspace}

\def\Bds{{\ensuremath{\B^0_{\left(\squark\right)}}}\xspace}

\newcommand{\SF}[2]{\ensuremath{#1 \times 10^{#2}}\xspace}

\newcommand{\SFSplitErr}[4]{\ensuremath{\left(#1 \pm #2 \pm #3 \right) \times 10^{#4}}\xspace}
\newcommand{\Err}[2]{\ensuremath{#1 \pm #2}\xspace}

\newcommand{\qSqrMuMu}{\ensuremath{\qsq\!\left( \mumu \right)}\xspace}

\newcommand{\Apm}{{\ensuremath{{\mathcal{A}}^{\pm\pm}}}\xspace}
\newcommand{\App}{{\ensuremath{{\mathcal{A}}^{++}}}\xspace}
\newcommand{\Amm}{{\ensuremath{{\mathcal{A}}^{--}}}\xspace}
\newcommand{\Azz}{{\ensuremath{{\mathcal{A}}^{00}}}\xspace}

\newcommand{\GenDecay}{\decay{\B}{\D\mumu}}
\newcommand{\GenDecays}{\B\,\to\D\mumu}
\newcommand{\GenDecayJpsi}{\decay{\B}{\D\jpsi}}
\newcommand{\GenDecayWS}{\decay{\B}{\D\mup\mup}}

\newcommand{\BdDecay}{\decay{\Bd}{\Dzb\mumu}}

\newcommand{\BdDecayJpsi}{\decay{\Bd}{\Dzb\jpsi}}

\newcommand{\BdDecayCharmless}{\decay{\Bd}{\jpsi\Kp\pim}}

\newcommand{\BdDecayLL}{\decay{\Bd}{\Dzb\ellell}}
\newcommand{\BdDecayRad}{\decay{\Bd}{\Dstarzb\gamma}}

\newcommand{\BdsDecay}{\decay{\Bds}{\Dzb\mumu}}
\newcommand{\BsDecay}{\decay{\Bs}{\Dzb\mumu}}

\newcommand{\BsDecayJpsi}{\decay{\Bs}{\Dzb\jpsi}}
\newcommand{\BdsDecayJpsi}{\decay{\Bds}{\Dzb\jpsi}}

\newcommand{\BuDecay}{\decay{\Bu}{\Dsp\mumu}}

\newcommand{\BuDecayJpsi}{\decay{\Bu}{\Dsp\jpsi}}

\newcommand{\BuDecayCharmless}{\decay{\Bp}{\jpsi\Kp\pim\pip}}

\newcommand{\BcDecay}{\decay{\Bc}{\Dsp\mumu}}

\newcommand{\BcDecayJpsi}{\decay{\Bc}{\Dsp\jpsi}}
\newcommand{\BcuDecay}{\decay{B^+_{(c)}}{\Dsp\mumu}}
\newcommand{\BcuDecayJpsi}{\decay{B^+_{(c)}}{\Dsp\jpsi}}

\newcommand{\BcDecayJpsiellpnu}{\mbox{\decay{\Bc}{\jpsi \ellp\nu}}}
\newcommand{\BcDecayJpsipip}{\mbox{\decay{\Bc}{\jpsi \pi^+}}}

\newcommand{\BcDecayDK}{\decay{\Bc}{\Dz\Kp}}

\newcommand{\BdDecaySLBkg}{\decay{\Bd}{\Dzb\pim\mup\neumb}}
\newcommand{\BdDecayDstrSLBkg}{\decay{\Bd}{\Dstarzb\pim\mup\neumb}}

\newcommand{\DstrzDecay}{\decay{\Dstarz}{\Dz\left(\piz/\g\right)}}

\newcommand{\BdDecayNorm}{\decay{\Bd}{\jpsi\Kstarz}}

\newcommand{\BdDecayNormLong} {\Bd \to \jpsi  \left( \to \mumu\right) \Kstarz \left( \to  \Kp\pim \right)}
\newcommand{\BuDecayNorm}{\decay{\Bu}{\jpsi\Kp}}

\newcommand{\BcDecayPartReco}{\decay{\Bc}{\Dssp\jpsi}}
\newcommand{\DsDecayDgamma}{\decay{\Dssp}{\Ds\gamma}}
\newcommand{\DsDecaypiz}{\decay{\Dssp}{\Ds\piz}}

\newcommand{\dbarkpicf}{\decay{\Dzb}{\Kp\pim}}

\newcommand{\dskkp}{\decay{\Dsp}{\Kp\Km\pip}}

\newcommand{\jpsimumu}{\decay{\jpsi}{\mumu}}

\usetikzlibrary{decorations.pathreplacing,decorations.pathmorphing,decorations.markings,trees,calc,patterns,snakes}
\tikzset{
photon/.style={decorate, decoration={snake}, draw=red},
line/.style={draw=blue},
particle/.style={draw=blue, postaction={decorate},decoration={markings,mark=at position .5 with {\arrow[draw=blue]{>}}}},
antiparticle/.style={draw=blue, postaction={decorate},decoration={markings,mark=at position .5 with {\arrow[draw=blue]{<}}}}, 
gluon/.style={decorate, draw=black,decoration={snake,amplitude=4pt, segment length=5pt}}, 
majorana/.style={draw=black, postaction={decorate},decoration={markings,mark=at position .48 with {\arrow[draw=black]{>}},mark=at position .52 with {\arrow[draw=black]{<}}}},
gluonloop/.style={circle, decorate, draw=black, decoration={coil,aspect=1.2,amplitude=2pt, segment length=4pt},minimum height=1.2em},
}

\begin{document}
\renewcommand{\thefootnote}{\fnsymbol{footnote}}
\setcounter{footnote}{1}


\begin{titlepage}
\pagenumbering{roman}

\vspace*{-1.5cm}
\centerline{\large EUROPEAN ORGANIZATION FOR NUCLEAR RESEARCH (CERN)}
\vspace*{1.5cm}
\noindent
\begin{tabular*}{\linewidth}{lc@{\extracolsep{\fill}}r@{\extracolsep{0pt}}}
\ifthenelse{\boolean{pdflatex}}
{\vspace*{-1.5cm}\mbox{\!\!\!\includegraphics[width=.14\textwidth]{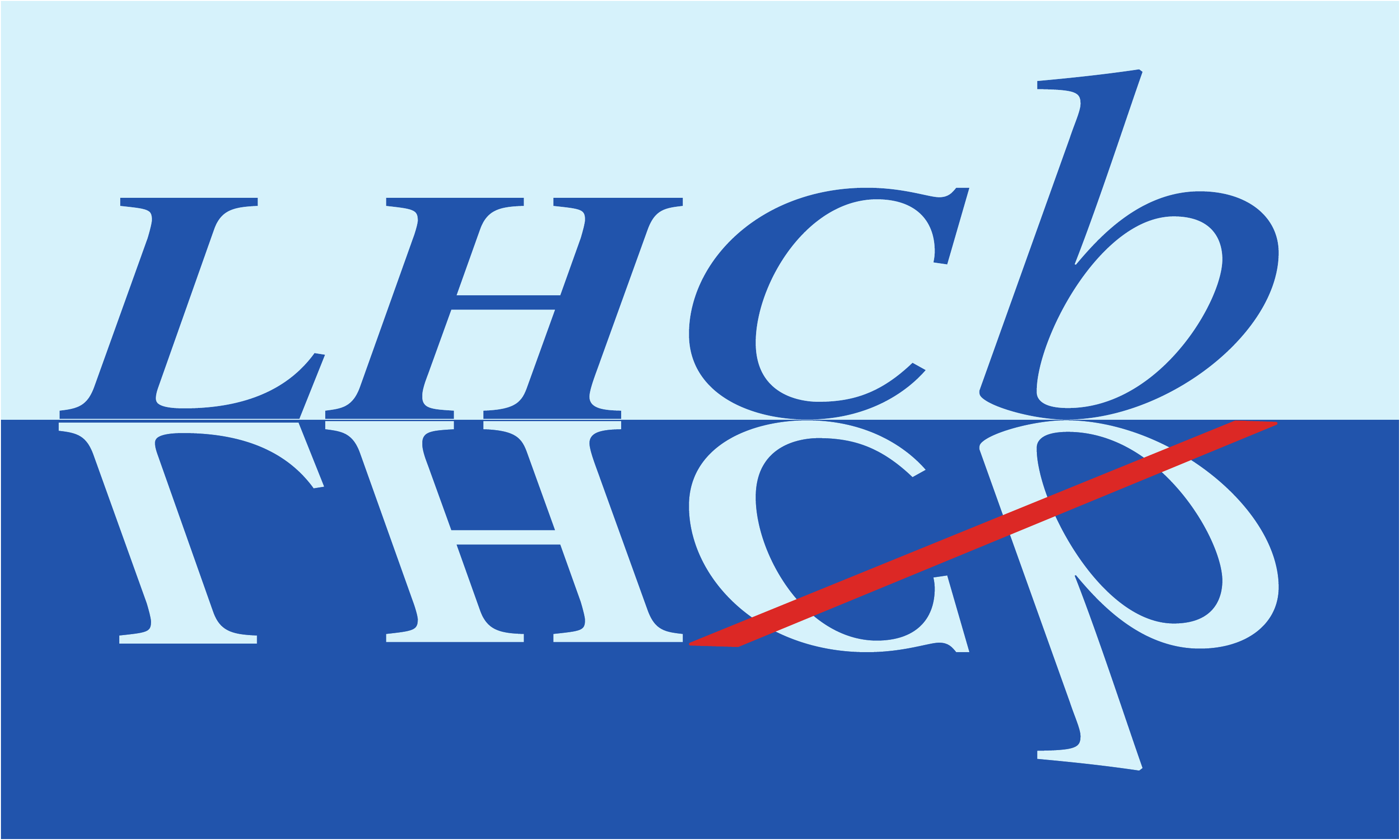}} & &}%
{\vspace*{-1.2cm}\mbox{\!\!\!\includegraphics[width=.12\textwidth]{figs/lhcb-logo.eps}} & &}%
\\
 & & CERN-EP-2023-121 \\  
 & & LHCb-PAPER-2022-048 \\  
 & & 26 February 2024 \\ 
 & & \\
\end{tabular*}

\vspace*{3.8cm}

{\normalfont\bfseries\boldmath\huge
\begin{center}
  \papertitle 
\end{center}
}

\vspace*{2.0cm}

\begin{center}
\paperauthors\footnote{Authors are listed at the end of this paper.}
\end{center}

\vspace{\fill}

\begin{abstract}
  \noindent
  A search for rare \GenDecay decays is performed using proton-proton collision data collected by the \lhcb experiment,  corresponding to an integrated luminosity of 9\invfb.
  No significant signals are observed in the non-resonant $\mu^+\mu^-$ modes, and upper limits of $\BraRat{\BdDecay}< \SF{5.1}{-8}$, $\BraRat{\BuDecay} < \SF{3.2}{-8}$, $\BraRat{\BsDecay} < \SF{1.6}{-7}$ and ${f_{c}}/{f_{u}}\cdot\BraRat{\BcDecay} < \SF{9.6}{-8}$ are set at the 95\,\% confidence level, where ${f_{c}}$ and ${f_{u}}$ are the fragmentation fractions of a  \B meson  with a $c$ and $u$ quark respectively in proton-proton collisions. Each  result is either the first such measurement or an improvement by three orders of magnitude on an existing limit. Separate upper limits are calculated when the muon pair originates from a \jpsimumu decay.  The branching fraction  of \BcDecayJpsi multiplied by  the fragmentation-fraction ratio  is measured to be
  \begin{equation*}
    \frac{f_{c}}{f_{u}}\cdot\mathcal{B}\left(\BcDecayJpsi\right) = \SFSplitErr{1.63}{0.15}{0.13}{-5}\,,
  \end{equation*}
  where the first uncertainty is statistical and the second systematic.
\end{abstract}

\vspace*{1.9cm}

\begin{center}
Published in JHEP 02 (2024) 032
\end{center}

\vspace{\fill}

{\footnotesize 
\centerline{\copyright~\papercopyright. \href{\paperlicenceurl}{\paperlicence}.}}
\vspace*{2mm}

\end{titlepage}


\newpage
\setcounter{page}{2}
\mbox{~}


\renewcommand{\thefootnote}{\arabic{footnote}}
\setcounter{footnote}{0}

\cleardoublepage


\pagestyle{plain} 
\setcounter{page}{1}
\pagenumbering{arabic}


\section{Introduction}
\label{sec:Introduction}

Heavy-flavour decays that are suppressed according to the Standard Model (SM) are probes for new particle currents with mass scales beyond the reach of direct searches. 
Recent work focuses on the measurement of partial rates in $b\to s\ell\ell$ transitions ($\ell\ell=\mu\mu$ or $ee$)~\cite{LHCb-PAPER-2015-051,LHCb-PAPER-2020-020,LHCb-PAPER-2020-041}.
Such flavour-changing neutral currents mediate $B\to K^{(*)}\mup\mun$ decays, predominantly via electroweak penguin amplitudes. The weak  amplitude  whereby the quark and antiquark of the initial meson undergo annihilation can also contribute, however this contribution can often be neglected due to suppression by a factor $\mathcal{O}(\Lambda_{QCD}/m_b)$~\cite{Beneke:2001at}. In order to validate this assumption, it is valuable to gather information about decays that are dominated by different modes of interaction between the quarks of the initial-state meson, either through a direct $b\to q\ell\ell$ decay, or proceeding through a $J/\psi\to\ell\ell $ intermediate state. 

The \BdsDecay decays{\footnote {Throughout this paper, the inclusion of charge-conjugate states is implied.}} 
proceed via an internal scatter process, specifically a \W-exchange as shown in \fig{fig:FeynmanDiagrams}(a). Similar \W-exchange decays, \BdsDecayJpsi, are depicted in \fig{fig:FeynmanDiagrams}(d). The clean experimental signature \jpsimumu can be used to search for this decay by isolating a region in the dimuon spectrum around the known \jpsi mass. For this mode, the \babar collaboration set a limit on the branching fraction of $\BraRat{\BdDecayJpsi}<\SF{1.3}{-5}$ at the 90\,\% confidence level (CL)~\cite{BaBar_2005}, which is 2--3 orders of magnitude greater than the theoretical expectation of \mbox{$10^{-8}$--$10^{-7}$}~\cite{PhysRevD.65.037504}.
An operator product expansion is used in Ref. \cite{Evans:1999zc} to predict a branching fraction of \SF{\sim3}{-9} for \BdDecayLL decays, integrated over a range of four-momentum transfer
$q^2$ between 1.0--5.0\gevgevcccc.
Elsewhere, a calculation based on perturbative QCD suggests a  higher value, $\mathcal{O}(10^{-5})$~\cite{Kim:2011ps}, although the treatment of long-distance divergences in that calculation has been questioned~\cite{Descotes-Genon:2001rya}.
No experimental limit has been set for the short-distance \BdDecayLL mode, but limits exist for the related radiative mode $\BraRat{\BdDecayRad}<\SF{2.5}{-5}$ at the 90\,\% CL~\cite{BaBar:2005ktr}, still greater than the $\sim10^{-6}$ prediction~\cite{Cheng:1994qw}.

\begin{figure}[t]
\begin{center}
\begin{tikzpicture}[scale=0.85]
\node at (0,3) [label={[label distance=2mm]left:(a)}] {};
\coordinate (a) at (0,2); 
\coordinate (ac) at (1.25,1.8);
\coordinate (c) at (2.5,2);
\coordinate (b) at (0,1); 
\coordinate (bd) at (1.25,1.2); 
\coordinate (d) at (2.5,1); 
\coordinate (e) at (1,2); 
\coordinate (f) at (1.5,1.85); 
\coordinate (g) at (2,2); 
\coordinate (h) at (2.25,3); 
\coordinate (i) at (3.25,2.75); 
\coordinate (j) at (3.25,3.25); 
\draw[antiparticle] (a) -- (ac);
\draw[antiparticle] (ac) -- (c);
\draw[particle] (b) -- (bd);
\draw[particle] (bd) -- (d);
\draw[dashed,red] (ac) -- (bd);
\draw[photon] (f) -- (h);
\draw[particle] (h) -- (i);
\draw[antiparticle] (h) -- (j);
\node at ($(a)$) [label={[label distance=-1.5mm]left:$\overline b$}] {};
\node at ($(b)$) [label={[label distance=-1.5mm]left:$d(s)$}] {};
\node at ($(c)$) [label={[label distance=-1.5mm]right:$\overline c$}] {};
\node at ($(d)$) [label={[label distance=-1.5mm]right:$u$}] {};
\node at ($(f)$) [label={[label distance=3mm]above:$\gamma$}] {};
\node at ($(i)$) [label={[label distance=0mm]right:$\mun$}] {};
\node at ($(j)$) [label={[label distance=0mm]right:$\mup$}] {};
\draw [black,decorate,decoration={brace,amplitude=5pt},xshift=20pt,yshift=0pt]
  (2.5,2.2)  -- (2.5,0.8) node [black,midway,right=0pt,xshift=5pt] {\Dzb};
\end{tikzpicture}
\hspace{2mm}
\begin{tikzpicture}[scale=0.85]
\node at (0,3) [label={[label distance=2mm]left:(b)}] {};
\coordinate (a) at (0,2); 
\coordinate (ac) at (0.75,1.5);
\coordinate (c) at (2.5,2);
\coordinate (b) at (0,1); 
\coordinate (bd) at (1.75,1.5); 
\coordinate (d) at (2.5,1); 
\coordinate (e) at (1,2); 
\coordinate (f) at (0.65,1.55); 
\coordinate (g) at (2,2); 
\coordinate (h) at (2.25,3); 
\coordinate (i) at (3.25,2.75); 
\coordinate (j) at (3.25,3.25); 
\draw[antiparticle] (a) -- (ac);
\draw[antiparticle] (bd) -- (c);
\draw[particle] (b) -- (ac);
\draw[particle] (bd) -- (d);
\draw[dashed,red] (ac) -- (bd);
\draw[photon] (f) -- (h);
\draw[particle] (h) -- (i);
\draw[antiparticle] (h) -- (j);
\node at ($(a)$) [label={[label distance=-1.5mm]left:$\overline b$}] {};
\node at ($(b)$) [label={[label distance=-1.5mm]left:$c(u)$}] {};
\node at ($(c)$) [label={[label distance=-1.5mm]right:$\overline s$}] {};
\node at ($(d)$) [label={[label distance=-1.5mm]right:$c$}] {};
\node at ($(f)$) [label={[label distance=1mm]above:$\gamma$}] {};
\node at ($(i)$) [label={[label distance=0mm]right:$\mun$}] {};
\node at ($(j)$) [label={[label distance=0mm]right:$\mup$}] {};
\draw [black,decorate,decoration={brace,amplitude=5pt},xshift=20pt,yshift=0pt]
  (2.5,2.2)  -- (2.5,0.8) node [black,midway,right=0pt,xshift=5pt] {\Dsp};
\end{tikzpicture}
\hspace{2mm}
\begin{tikzpicture}[scale=0.85]
\node at (0,3) [label={[label distance=2mm]left:(c)}] {};
\coordinate (a) at (0,2); 
\coordinate (ac) at (0.75,2);
\coordinate (c) at (2.5,2);
\coordinate (b) at (0,1); 
\coordinate (bd) at (1.75,2); 
\coordinate (d) at (2.5,1); 
\coordinate (e) at (1,2.5); 
\coordinate (f) at (1.5,2.25); 
\coordinate (g) at (2,2); 
\coordinate (h) at (2.25,3); 
\coordinate (i) at (3.25,2.75); 
\coordinate (j) at (3.25,3.25); 
\draw[antiparticle] (a) -- (ac);
\draw[antiparticle] (bd) -- (c);
\draw[antiparticle] (ac) to [in=90,out=90] (bd);
\draw[dashed,red] (ac) to [in=270,out=270] (bd);
\draw[particle] (b) -- (d);
\draw[photon] (f) -- (h);
\draw[particle] (h) -- (i);
\draw[antiparticle] (h) -- (j);
\node at ($(a)$) [label={[label distance=-1.5mm]left:$\overline b$}] {};
\node at ($(b)$) [label={[label distance=-1.5mm]left:$c$}] {};
\node at ($(c)$) [label={[label distance=-1.5mm]right:$\overline  s$}] {};
\node at ($(d)$) [label={[label distance=-1.5mm]right:$c$}] {};
\node at ($(f)$) [label={[label distance=1mm]above:$\gamma$}] {};
\node at ($(i)$) [label={[label distance=0mm]right:$\mun$}] {};
\node at ($(j)$) [label={[label distance=0mm]right:$\mup$}] {};
\draw [black,decorate,decoration={brace,amplitude=5pt},xshift=20pt,yshift=0pt]
  (2.5,2.2)  -- (2.5,0.8) node [black,midway,right=0pt,xshift=5pt] {\Dsp};
\end{tikzpicture}

\vspace{3mm}
\begin{tikzpicture}[scale=0.85]
\node at (0,3.25) [label={[label distance=2mm]left:(d)}] {};
\coordinate (a) at (0,3); 
\coordinate (ac) at (1.25,2.8);
\coordinate (c) at (2.5,3.0);
\coordinate (b) at (0,1); 
\coordinate (bd) at (1.25,1.2); 
\coordinate (d) at (2.5,1); 
\coordinate (e) at (1,2); 
\coordinate (f) at (1.5,2.85); 
\coordinate (g) at (2.5,2); 
\coordinate (h) at (2,2); 
\coordinate (i) at (2.5,2.5); 
\coordinate (j) at (2.5,1.5); 
\draw[antiparticle] (a) -- (ac);
\draw[antiparticle] (ac) -- (c);
\draw[particle] (b) -- (bd);
\draw[particle] (bd) -- (d);
\draw[dashed,red] (ac) -- (bd);
\draw[particle] (h) to [in=180,out=90] (i);
\draw[antiparticle] (h) to [in=180,out=270] (j);
\node at ($(a)$) [label={[label distance=-1.5mm]left:$\overline b$}] {};
\node at ($(b)$) [label={[label distance=-1.5mm]left:$d(s)$}] {};
\node at ($(c)$) [label={[label distance=-1.5mm]right:$\overline c$}] {};
\node at ($(d)$) [label={[label distance=-1.5mm]right:$u$}] {};
\node at ($(i)$) [label={[label distance=-1.5mm]right:$c$}] {};
\node at ($(j)$) [label={[label distance=-1.5mm]right:$\overline c$}] {};
\draw [black,decorate,decoration={brace,amplitude=5pt},xshift=20pt,yshift=0pt]
  (2.5,3.1)  -- (2.5,2.3) node [black,midway,right=0pt,xshift=5pt] {\jpsi};
\draw [black,decorate,decoration={brace,amplitude=5pt},xshift=20pt,yshift=0pt]
  (2.5,1.6)  -- (2.5,0.8) node [black,midway,right=0pt,xshift=5pt] {\Dzb};
\end{tikzpicture}
\hspace{2mm}
\begin{tikzpicture}[scale=0.85]
\node at (0,3.25) [label={[label distance=2mm]left:(e)}] {};
\coordinate (a) at (0,3); 
\coordinate (ac) at (0.75,2);
\coordinate (c) at (2.5,3);
\coordinate (b) at (0,1); 
\coordinate (bd) at (1.5,2); 
\coordinate (d) at (2.5,1); 
\coordinate (e) at (1,2); 
\coordinate (f) at (1.5,2.85); 
\coordinate (g) at (2.5,2); 
\coordinate (h) at (2,2); 
\coordinate (i) at (2.5,2.5); 
\coordinate (j) at (2.5,1.5); 
\draw[antiparticle] (a) -- (ac);
\draw[particle] (bd) -- (c);
\draw[particle] (b) -- (ac);
\draw[antiparticle] (bd) -- (d);
\draw[dashed,red] (ac) -- (bd);
\draw[antiparticle] (h) to [in=180,out=90] (i);
\draw[particle] (h) to [in=180,out=270] (j);
\node at ($(a)$) [label={[label distance=-1.5mm]left:$\overline b$}] {};
\node at ($(b)$) [label={[label distance=-1.5mm]left:$c(u)$}] {};
\node at ($(c)$) [label={[label distance=-1.5mm]right:$c$}] {};
\node at ($(d)$) [label={[label distance=-1.5mm]right:$\overline s$}] {};
\node at ($(i)$) [label={[label distance=-1.5mm]right:$\overline c$}] {};
\node at ($(j)$) [label={[label distance=-1.5mm]right:$c$}] {};
\draw [black,decorate,decoration={brace,amplitude=5pt},xshift=20pt,yshift=0pt]
  (2.5,3.1)  -- (2.5,2.3) node [black,midway,right=0pt,xshift=5pt] {\jpsi};
\draw [black,decorate,decoration={brace,amplitude=5pt},xshift=20pt,yshift=0pt]
  (2.5,1.6)  -- (2.5,0.8) node [black,midway,right=0pt,xshift=5pt] {\Dsp};
\end{tikzpicture}
\hspace{2mm}
\begin{tikzpicture}[scale=0.85]
\node at (0,3.25) [label={[label distance=2mm]left:(f)}] {};
\coordinate (a) at (0,3); 
\coordinate (ac) at (1,3);
\coordinate (c) at (2.5,3);
\coordinate (b) at (0,1); 
\coordinate (bd) at (1.75,2); 
\coordinate (d) at (2.5,1); 
\coordinate (e) at (1,2.5); 
\coordinate (f) at (1,2.25); 
\coordinate (g) at (2,2); 
\coordinate (h) at (1.75,2); 
\coordinate (i) at (2.5,2.5); 
\coordinate (j) at (2.5,1.5); 
\draw[antiparticle] (a) -- (c);
\draw[dashed,red] (ac) -- (bd);
\draw[particle] (b) -- (d);
\draw[particle] (h) -- (i);
\draw[antiparticle] (h) -- (j);
\node at ($(a)$) [label={[label distance=-1.5mm]left:$\overline b$}] {};
\node at ($(b)$) [label={[label distance=-1.5mm]left:$c$}] {};
\node at ($(c)$) [label={[label distance=-1.5mm]right:$\overline  c$}] {};
\node at ($(d)$) [label={[label distance=-1.5mm]right:$c$}] {};
\node at ($(i)$) [label={[label distance=-1.5mm]right:$c$}] {};
\node at ($(j)$) [label={[label distance=-1.5mm]right:$\overline s$}] {};
\draw [black,decorate,decoration={brace,amplitude=5pt},xshift=20pt,yshift=0pt]
  (2.5,3.1)  -- (2.5,2.3) node [black,midway,right=0pt,xshift=5pt] {\jpsi};
\draw [black,decorate,decoration={brace,amplitude=5pt},xshift=20pt,yshift=0pt]
  (2.5,1.6)  -- (2.5,0.8) node [black,midway,right=0pt,xshift=5pt] {\Dsp};
\end{tikzpicture}
\end{center}
  \caption{
    \small 
    The leading diagrams for the decays: (a) \BdsDecay , (b) \BcuDecay , (c)~\BcDecay , (d) \BdsDecayJpsi , (e) \BcuDecayJpsi and (f) \BcDecayJpsi. Dashed lines represent the charged weak current, and a gluon exchange with unconnected quark lines is implied. Where two quark currents are indicated, the quark in parenthesis leads to a CKM-suppressed process.
    }
  \label{fig:FeynmanDiagrams}
\end{figure}

The present analysis is extended to include charged \GenDecay decays. Knowledge of the decays of doubly-heavy \Bc mesons is accumulating rapidly, though most observed decays can be attributed to favoured amplitudes  with an external $W$-emission, \eg~\BcDecayJpsiellpnu ~and \BcDecayJpsipip
decays~\cite{CDF:1998ihx,LHCB-PAPER-2014-050}. The \BcDecayJpsi decay, first reported in Ref.~\cite{LHCb-PAPER-2013-010} and measured more recently in Refs.~\cite{ATLAS_2016_Bc2DsJpsi, ATLAS_2022_Bc2DsJpsi}, is mediated by the external $W$-emission process but is also sensitive to colour-suppressed, annihilation and electroweak penguin amplitudes (\fig{fig:FeynmanDiagrams}(f), (e) and (c), respectively). 
To assess the relative magnitude of these sub-leading amplitudes, measurements across a range of suppressed decays must be made. The partial decay rate of the recently observed \BcDecayDK decay together with the absence of $\Bc\to D^0\pi^+$ decays~\cite{LHCB-PAPER-2016-058} implies the former is dominated by annihilation or electroweak penguin contributions. Charmless \Bc decays isolate the annihilation diagram, but no evidence for such decays has yet emerged~\cite{LHCb-PAPER-2013-034, LHCB-PAPER-2016-001}. This paper presents the search for  yet-unobserved \BcDecay decays, which are sensitive to  electroweak penguin and radiative annihilation transitions only (\fig{fig:FeynmanDiagrams}(b) and (c), respectively). 
The SM branching fraction for this decay mode is predicted to be $\mathcal{O}(10^{-8})$ but contributions from beyond-SM processes can raise this by an order of magnitude~\cite{Maji:2020wer}.
Similarly, since the \Dsp meson in the \BuDecay decay does not maintain any of the initial-state quark content, this transition must proceed via annihilation (\fig{fig:FeynmanDiagrams}(b) and (e)). No search results have yet been reported for such decays.

The  \GenDecay decays depicted in \fig{fig:FeynmanDiagrams} represent eight independent decay modes whose search is reported in this paper, namely:
\begin{itemize}
\item $\Bd$ decay modes: $\BdDecay$, $\BdDecayJpsi$,
\item $\Bu$ decay modes: $\BuDecay$, $\BuDecayJpsi$,
\item $\Bs$ decay modes: $\BsDecay$, $\BsDecayJpsi$,
\item $\Bc$ decay modes: $\BcDecay$, $\BcDecayJpsi$.
\end{itemize}
The decay modes $B^+\to D^+\mu^+\mu^-$ and $B_c^+\to D^+\mu^+\mu^-$, either proceeding directly to $\mu^+\mu^-$ or through a   $J/\psi$ intermediate state,  are not included here.  There are large Cabbibo suppression factors associated with these modes with respect to the \BuDecay and \BcDecay 
 modes, and as such are significantly below our current sensitivity.

The searches are performed using proton-proton collision data collected by the \lhcb experiment  at centre-of-mass energies of 7, 8 and 13\tev, corresponding to a total integrated luminosity of 9\invfb. 
The \Dzb meson candidates are reconstructed in the \dbarkpicf channel,  and the \Ds meson candidates in the \dskkp channel. 
The \jpsi candidates are reconstructed in the \jpsimumu decay mode.
For the non-\jpsi modes, the square of the dimuon invariant mass, \qSqrMuMu ,  is selected within the range 0.044--8.0\gevgevcccc , which excludes the \jpsi region.
The selection, efficiency and background corrections are  finalised before the signal measurements are performed, following a strategy of blinding to ensure unbiased results.

For normalisation, kinematically similar decays with higher branching fractions are also reconstructed. The $\BdDecayNormLong$ decay is used to normalise the neutral \Bz and \Bs signal modes, while the \BuDecayNorm decay is used to normalise the two \Bu signal modes and the \BcDecayJpsi measurement. The  \BcDecay decay is normalised with respect to the observed \BcDecayJpsi decay mode.

\section{Detector, triggering and simulation}
\label{sec:Detector}

The \lhcb detector~\cite{LHCb-DP-2008-001,LHCb-DP-2014-002} is a single-arm forward
spectrometer covering the \mbox{pseudorapidity} range $2<\eta <5$,
and designed for the study of particles containing \bquark or \cquark
quarks. The detector includes a high-precision tracking system
consisting of a silicon-strip vertex detector (VELO)~\cite{LHCb-DP-2014-001} surrounding the proton-proton
interaction region, a large-area silicon-strip detector~\cite{LHCb-DP-2013-003}  located
upstream of a dipole magnet with a bending power of about
$4{\mathrm{\,Tm}}$, and three stations of silicon-strip detectors and straw
drift tubes 
\cite{LHCb-DP-2017-001}
placed downstream of the magnet.
The tracking system provides a measurement of the momentum, \ptot, of charged particles with
a relative uncertainty that varies from 0.5\,\% at low momentum to 1.0\,\% at 200\gevc.
The minimum distance of a track to a primary vertex, the impact parameter (IP), 
is measured with a resolution of $(15+29/\pt)\mum$,
where \pt is the component of the momentum transverse to the beam, expressed in\,\gevc.
Different types of charged hadrons are distinguished using information
from two ring-imaging Cherenkov (RICH) detectors~\cite{LHCb-DP-2012-003}. 
Photons, electrons and hadrons are identified by a calorimeter system consisting of
scintillating-pad and preshower detectors, an electromagnetic
and a hadronic calorimeter. Muons are identified by a
system composed of alternating layers of iron and multiwire
proportional chambers~\cite{LHCb-DP-2012-002}.

The online event selection is performed by a trigger \cite{LHCb-DP-2012-004, Aaij_2019-trigger}, consisting of a hardware stage, based on information from the calorimeter and muon systems, followed by a software stage that performs a full event reconstruction. The hardware trigger selects \GenDecay candidates containing at least one muon with a large transverse momentum, a pair of muons with a large product of their transverse momenta, or containing large energy deposits in the calorimeters from particles related to the \D-meson decay. The transverse momentum threshold of the muon triggers varied in the range between 1 and 3\gevc, depending on the data-taking conditions, and similarly the hadronic energy threshold varied between 3 and 4\gevc. At the software stage, the trigger required a two-, three- or four-track secondary vertex with a significant displacement from any proton-proton interaction vertex \cite{BBDT}.
At least one charged particle must have a \pt  greater than 1.5\gevc.

Simulation is used to measure the efficiency of both the detector acceptance and the applied selection requirements.
In the simulation, proton-proton collisions are generated using
\pythia~\cite{Sjostrand:2007gs, Sjostrand:2006za} 
with a specific \lhcb configuration~\cite{LHCb-PROC-2010-056}.
Unstable particles
are described by \evtgen~\cite{Lange:2001uf}, with the rare-$\B$ modes decayed uniformly in phase space and final-state radiation generated using \photos~\cite{Golonka:2005pn}.
The interaction of the generated particles with the detector, and its response,
are implemented using the \geant
toolkit~\cite{Allison:2006ve, Agostinelli:2002hh} as described in
Ref.~\cite{LHCb-PROC-2011-006}.

\section{Candidate selection}
\label{sec:Selection}

When reconstructing  \D meson candidates, all hadron tracks are required to be compatible with their respective particle hypotheses according to a likelihood-based variable that uses mainly the RICH input from the particle identification (PID) system. The candidates must have a good-quality vertex and  have travelled a significant distance from the \B -candidate decay vertex. The $\Dzb\to\Kp\pim$ ($\Ds\to K^+K^-\pip$) combinations are required to be within  $\pm 25$\mevcc ($\pm 20$\mevcc ) of the known \Dzb (\Ds ) mass~\cite{PDG2022},  corresponding to roughly $\pm 3$ standard-deviation windows around the respective mass peaks.

A PID cut is applied to both muon candidates,  using an equivalent likelihood-based
variable type as for the hadrons. For the signal modes involving a \jpsimumu reconstruction, the dimuon invariant mass is required to be within $\pm 36$\mevcc of the known \jpsi mass~\cite{PDG2022}, \ie  a $\pm 3$ standard-deviation window.  For the non-\jpsi signal decays,  \qSqrMuMu    is selected within the range 0.044--8.0\gevgevcccc.

With the \D meson and muon pair identified, \B meson candidates are then reconstructed with the \D , and where appropriate the \jpsi , masses constrained to their known values~\cite{Hulsbergen:2005pu}. For modes with a primary \Bz, \Bs or \Bp meson, a decay time greater than 0.2\ps is required, whereas for modes with a primary \Bc meson, the minimum decay time is relaxed to 0.05\ps due to the shorter lifetime of this particle.
These values can be compared to the measured lifetime resolution at LHCb which is approximately 50\,fs~\cite{LHCb-DP-2014-001}.  
The  decay-time requirements correspond to typical flight paths of around 7\,mm for the \Bz, \Bs and \Bp and 2.5\,mm for the \Bc ,  to be compared to the flight-path resolution of approximately 230\,$\mu$m.

A boosted decision tree~(BDT) algorithm \cite{Breiman,AdaBoost}, implemented using the TMVA 
toolkit~\cite{Hocker:2007ht, TMVA4}, is used to separate signal from combinatorial background. The BDT classifier is trained and applied independently for  three families of modes:  neutral \B mesons including both \Bz and \Bs (since the final states for the searches are identical),  charged \B mesons, and  \Bc candidates.
Variables describing the properties  of the \B and \D meson candidates are provided to train the BDT classifiers: a total of 15 for the neutral decays and an additional two for \Bp and \Bc modes which include a \Ds decay. 
The variables used comprise kinematic and topological quantities
such as \pt and IP, 
flight distance, direction of the weakly decaying heavy mesons and the angle between their reconstructed trajectory and momentum.
Simulated samples containing signal decays are used as target samples in the BDT training, while background samples are taken from data. For the neutral \B meson BDT algorithm, decays with a \Bz invariant  mass greater than 5800\mevcc are used to form the background sample. For the charged \B meson BDT classifiers, a pure combinatorial background sample of  wrong-charge \GenDecayWS combinations is used; this is to avoid feed-down effects from \Bc decays if training on the \Bp upper mass sideband. 
The selection that is applied to the BDT output is optimised using a Punzi figure of merit~\cite{Punzi:2003bu} to choose a balance between signal efficiency and background rejection for each of the eight search categories.

The normalisation samples containing prolific \BdDecayNorm and \BuDecayNorm decays are selected without a multivariate technique. The \jpsi and \Kstarz candidates are required to  have a good-quality vertex and be within $\pm 50$\mevcc of their known masses. The particle types of their decay-product tracks are required to be correctly identified using PID selection criteria. Additional selection requirements are applied on the transverse momenta of the final-state tracks.
The yields of \BdDecayNorm and \BuDecayNorm decays are determined using binned maximum-likelihood fits to data. 
An S-wave contribution of 4\% under the \Kstarz resonance peak is numerically subtracted from the fitted \BdDecayNorm yield as for  Ref.~\cite{LHCb-PAPER-2013-023}. 

\section{Signal yield estimation}
\label{sec:Fits}

\begin{figure}[tb]
  \begin{center}
    \includegraphics[width=0.49\linewidth]
    {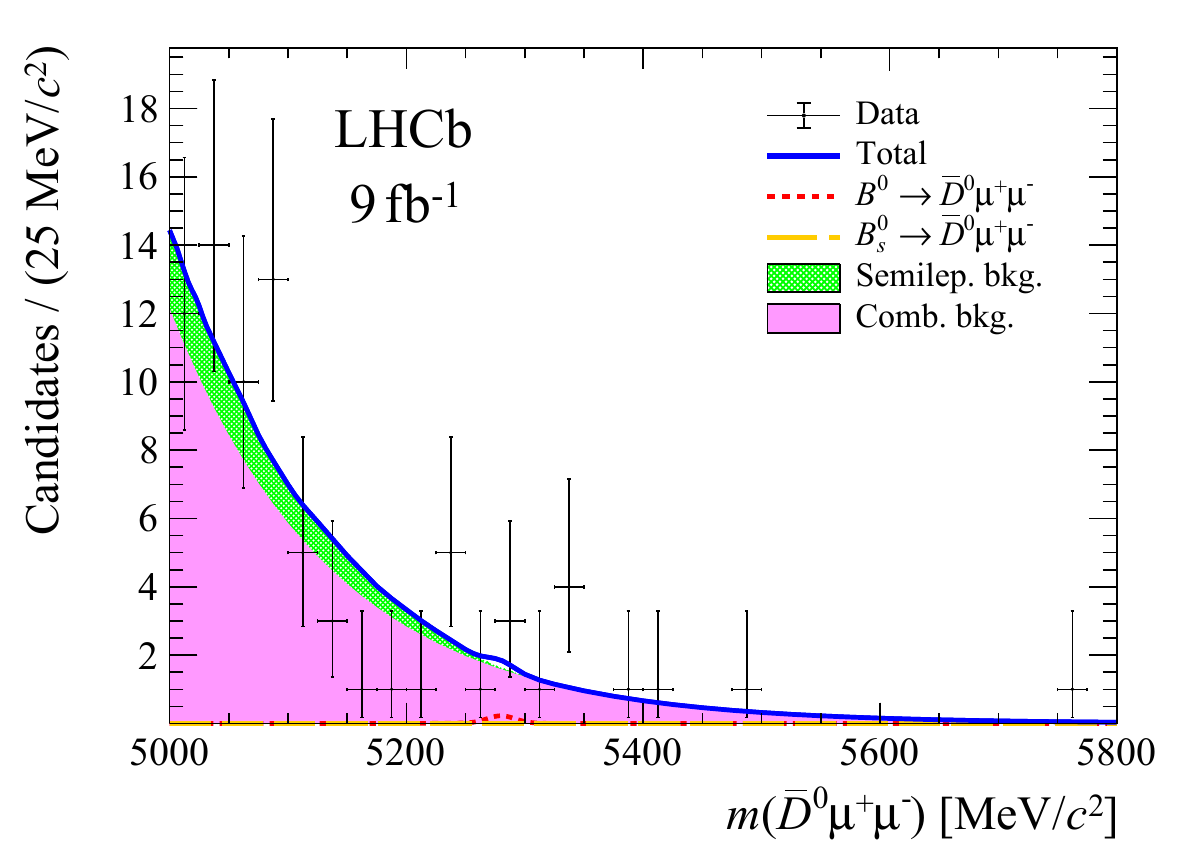}
    \includegraphics[width=0.49\linewidth]
    {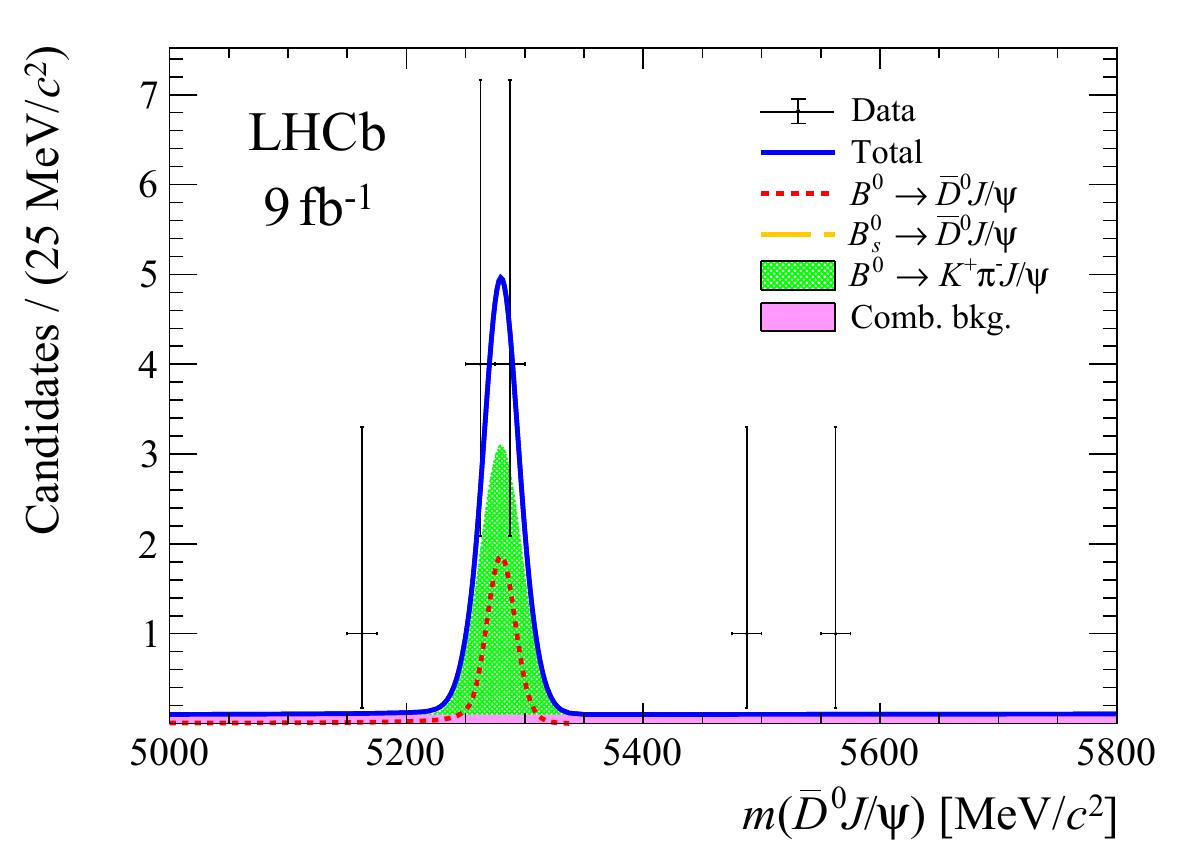}
    \vspace*{-0.5cm}
  \end{center}
  \caption{
    \small
	Invariant-mass distributions of (left) \BdsDecay candidates and (right) {\mbox \BdsDecayJpsi} candidates with the fits superimposed.
    }
  \label{fig:BdMassFit}
\end{figure}

\begin{figure}[htb]
  \begin{center}
    \includegraphics[width=0.49\linewidth]{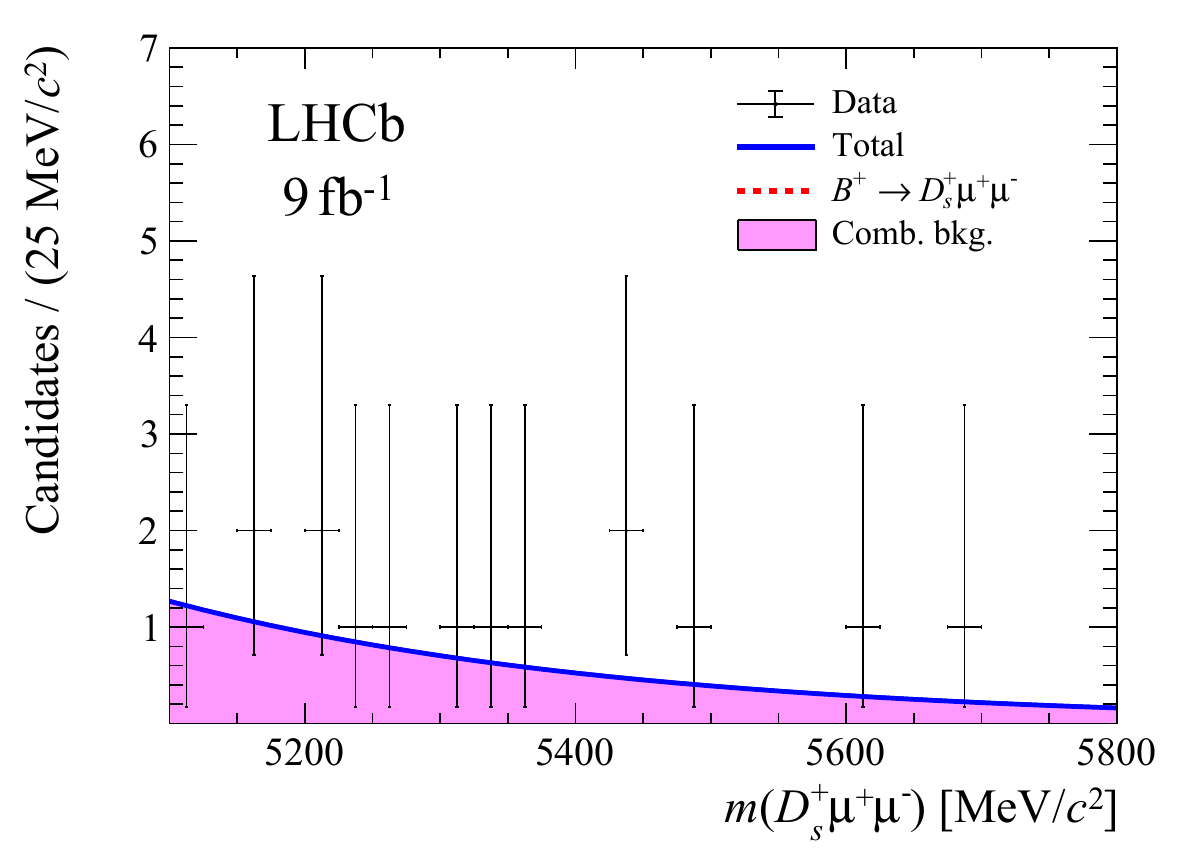}
    \includegraphics[width=0.49\linewidth]
    {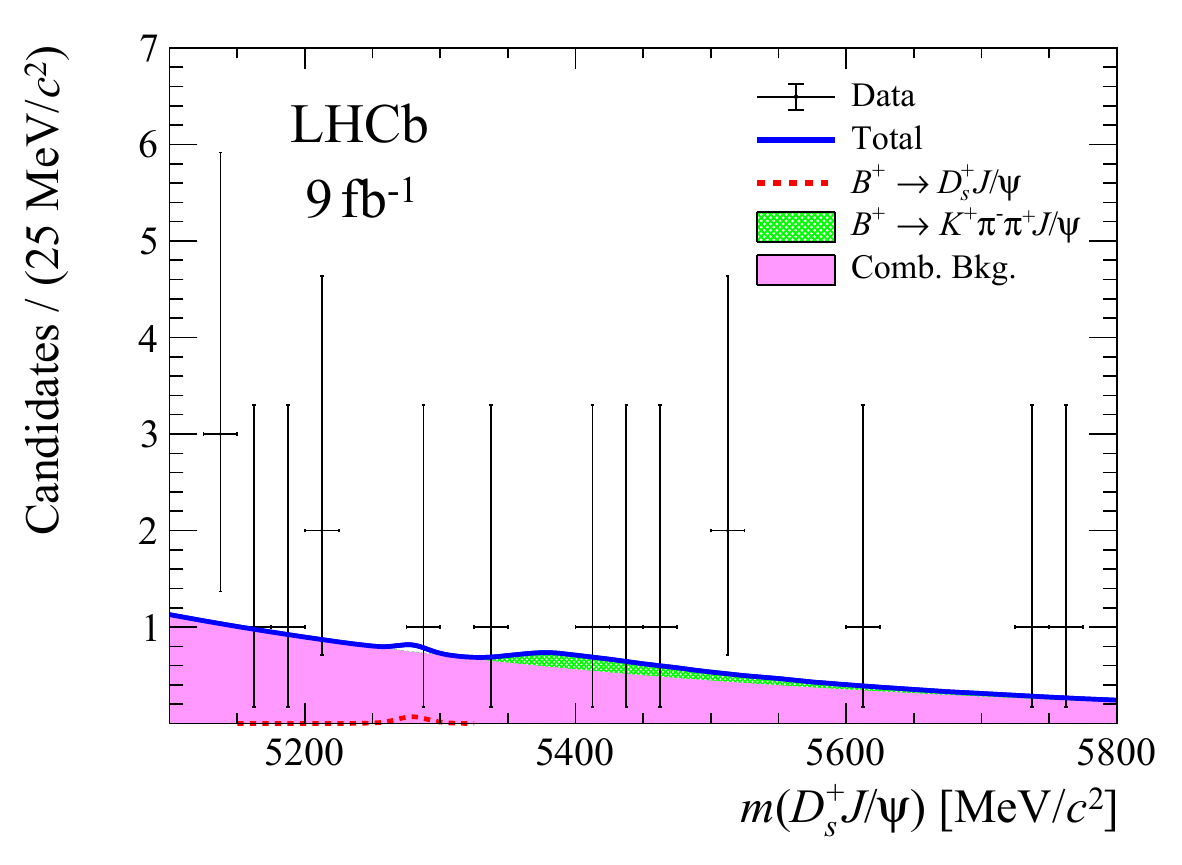}
    \vspace*{-0.5cm}
  \end{center}
  \caption{
    \small 
	Invariant-mass distributions of (left) \BuDecay candidates and (right) {\mbox \BuDecayJpsi}  candidates with the fits superimposed.
    }
  \label{fig:BuMassFit}
\end{figure}

\begin{figure}[tb]
  \begin{center}
    \includegraphics[width=0.49\linewidth]{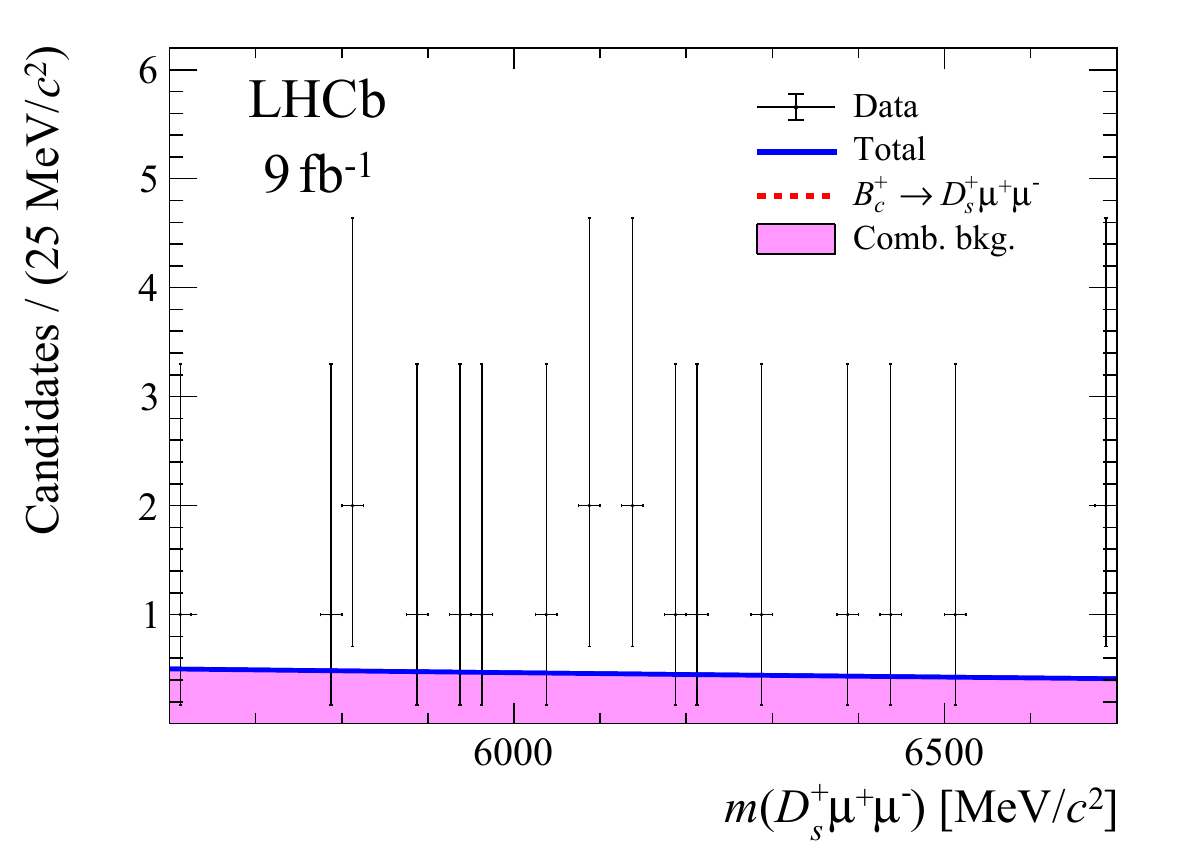}
    \includegraphics[width=0.49\linewidth]{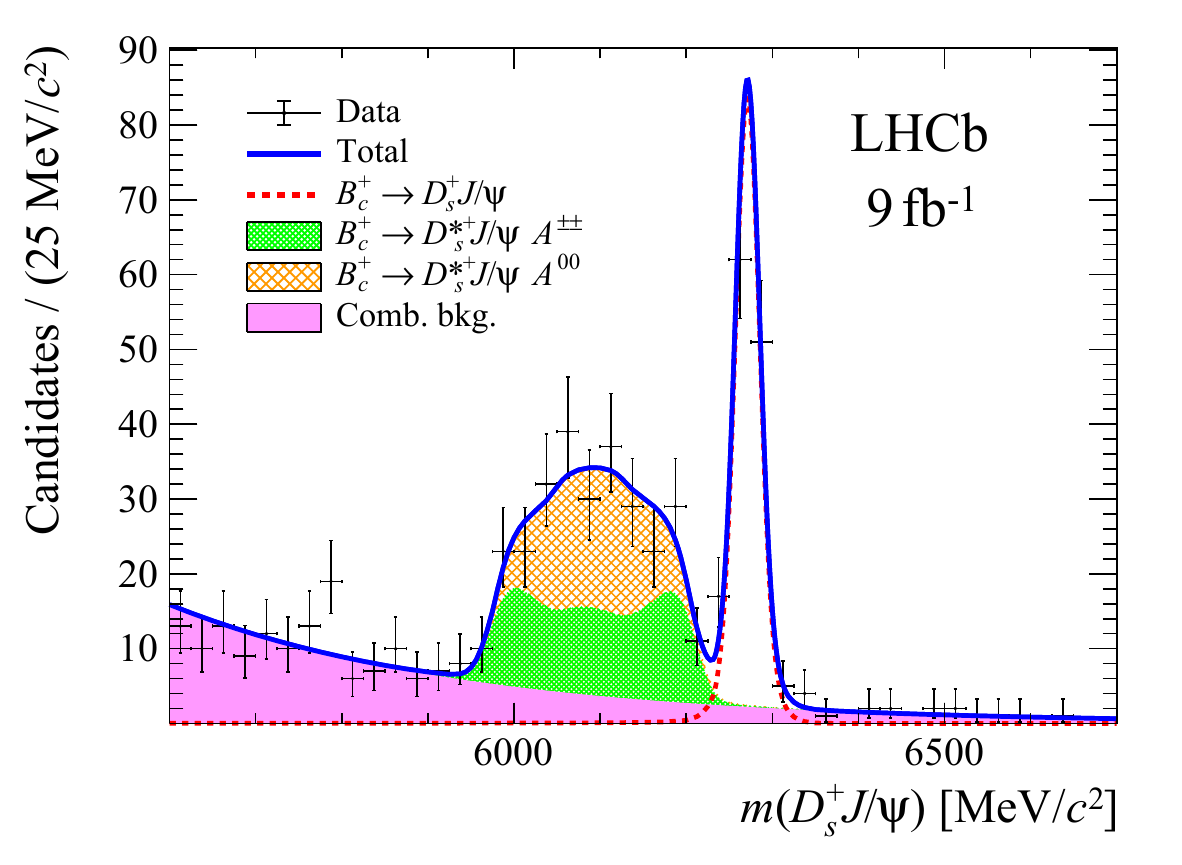}
    \vspace*{-0.5cm}
  \end{center}
  \caption{
    \small 
	Invariant-mass distributions of (left) \BcDecay candidates  and (right) {\mbox \BcDecayJpsi} candidates with the fits superimposed. 
    }
  \label{fig:BcMassFit}
\end{figure}

Six invariant-mass distributions of \B candidate decays are fitted, corresponding to \BdsDecay, \BuDecay  and \BcDecay, selected with and without a \jpsi intermediate state. 
In all samples, the distributions are described by probability density functions (PDFs) with shape parameters that are free to vary. The yields of signal and background contributions are measured using extended unbinned maximum-likelihood fits to the invariant-mass distributions, with the single exception of the \BdsDecayJpsi decay where the background contribution is fixed as described below.   Signal \GenDecay decays are modelled by the sum of two Crystal-Ball functions~\cite{Skwarnicki:1986xj}, each consisting of a Gaussian core and a power-law tail, where the widths and tail parameters are fixed based on simulation. In the case of the neutral \B invariant-mass distributions, two such functions are used to describe each of the $\B^0$ and $B^0_{s}$ candidate decays. All means are fixed from  values measured in the appropriate normalisation-mode fit or, for the \Bs modes where normalisation-mode  yields are lower, to their known values~\cite{PDG2022} with a small offset derived from simulation.
In all samples, the combinatorial backgrounds are modelled by exponential distributions with parameters that are free to vary. Specific background contributions are also included, which are described below.
The fits to the invariant-mass distributions for the \GenDecay and \GenDecayJpsi decays are shown in 
Figs.~\ref{fig:BdMassFit}, \ref{fig:BuMassFit} and \ref{fig:BcMassFit}
for neutral $B^0_{(s)}$, charged \Bp and \Bc mesons, respectively. The fit ranges are defined by the limits of the plots.  

It is necessary to include additional background components to fully describe the data for specific decay modes.
In the \BdDecay invariant-mass distribution, \BdDecaySLBkg and \BdDecayDstrSLBkg decays with \DstrzDecay constitute a background if the \pim is misidentified as a \mun and the neutral particles are not reconstructed. This contribution is modelled using a kernel density estimation technique~\cite{Cranmer_2001} on events from a simplified LHCb simulation~\cite{Cowan_2017}. These sources are combined into a single distribution using relative branching fractions and efficiencies.

For the invariant-mass fit of the \BdsDecayJpsi decay, 
non-resonant \BdDecayCharmless decays
are modelled with a single Gaussian distribution. The width is fixed to that of the signal peak and the yield is taken from fits to lower and upper $m(\Dz)$ sidebands, defined between the ranges 1785--1835\mevcc and 1900--1950\mevcc, respectively. Averaging between sidebands, the non-resonant contribution within the \Dz mass range is estimated to be \Err{5.2}{2.9} events, with the yield then fixed to this central value.
For the mass fit of the \BuDecayJpsi decay, 
\BuDecayCharmless decays may enter if the \pim is misidentified as a \Km.  The shape of such a background is determined using simulation and the yield is fixed in the same manner as for the background in the \Bz samples.

In the fits to the invariant-mass distributions of the $B_c$ modes show in  Fig.~\ref{fig:BcMassFit}, no {\mbox \BcDecay} signal is observed whilst a \BcDecayJpsi signal is clearly visible. For the \BcDecayJpsi candidates, \BcDecayPartReco decays are considered where \Dssp  decays into 
\DsDecayDgamma or \DsDecaypiz and 
 the neutral particles are not reconstructed. This  results in the broad structure seen between 5.9
and 6.2\gevcc in Fig.~\ref{fig:BcMassFit}. As \BcDecayPartReco is a  decay of a pseudoscalar into two vector particles, the decay is described by three helicity amplitudes: \Azz (longitudinal polarisation), \App and \Amm (transverse polarisations). The latter two are combined into a single \Apm PDF because the \Ds\jpsi invariant-mass distribution is identical for the \App and \Amm amplitudes.
The two helicity contributions are described by analytic PDFs in the fit to data.
The \Apm distribution
is parameterised by an upward-open parabola whose range is defined by the kinematic
endpoints of the decay and convolved with a Gaussian resolution function, resulting in a characteristic double-peaked shape. 
The \Azz   distribution is described by a parabola exhibiting a maximum  and convolved with a Gaussian resolution function, resulting in a broad single peak. These PDFs are further described in Ref.~\cite{LHCb-PAPER-2017-021}. All shape parameters are fixed from fits to simulation.

\section{Branching fraction determination}
\label{sec:BRDetermination}

The signal branching fractions are obtained from efficiency-corrected yield measurements of the signal and normalisation modes scaled by the known branching fractions. Specifically,
\begin{equation}
\label{eqn:BRCalc}
    \BraRat{\GenDecay} = 
    \frac{N^{\GenDecays}}{N^{\rm{norm}}} \cdot 
	\frac{\varepsilon^{\rm{norm}}}{\varepsilon^{\GenDecays}} \cdot 
	\frac{\mathcal{B}^{\rm{norm}}_{\rm{sub{\text -}decays}}}
    {\mathcal{B}_{\rm{sub{\text -}decays}}^{\GenDecays}} \cdot 
	\mathcal{B}^{\rm{norm}},
\end{equation}
where $\BraRat{\GenDecay}$ is the branching fraction of the relevant signal decay, $N^{i}$ refers to the observed yield for the signal decay or normalisation mode, $\varepsilon^{i}$ is the corresponding total selection efficiency, and $\mathcal{B}^{\rm{norm}}$ is the branching fraction of the normalisation channel. The product of branching fractions of the relevant sub-decays, $\mathcal{B}_{\rm{sub{\text -}decays}}^{\,i}$, are  listed with all 
relevant branching fractions  in Table~\ref{tab:Branchingratios}. These values are taken from Ref.~\cite{PDG2022}, with the exception of $\BraRat{\BcDecayJpsi}$ which uses the value measured in this paper when used as a normalisation mode for the \BcDecay decay.
For the \Bs modes, the result takes into account the ratio of the fragmentation fractions of \Bs to \Bd meson production at \lhcb, $f_{s}\big/f_{d}$. The value used in this paper is $\Err{0.249}{0.020}$,   taken from Ref.~\cite{LHCb-PAPER-2020-046}, which is a luminosity-weighted average taken over the values at the three centre-of-mass energies  7, 8 and 13\tev . The  quoted error  is associated with the averaging procedure, and also includes an uncertainty due to the integrated  \pt dependence of the $f_{s}\big/f_{d}$ values.
The analogous quantity for \Bc production,  
$f_{c}\big/f_{u}= (\Err{7.5}{1.8})\times10^{-3}$~\cite{LHCb-PAPER-2019-033}, is also luminosity averaged, and is kept as a separate factor in the quoted results for the two \Bc modes due to the relatively poor determination of this quantity to date.

\begin{table}[htb]
\centering
\caption{\small
A summary of the branching fractions of the decays (taken from Ref.~\cite{PDG2022}) and the fragmentation ratio (taken from Ref.~\cite{LHCb-PAPER-2020-046}) which are used in the calculation of  \eqinline{eqn:BRCalc}.  The value of  $\BraRat{\Kstarz\to\Kp\pim}$ is taken from the ratio of isospin amplitudes.}
\begin{tabular}{lc}
 Measurement  &  Value \\
\hline 
\vspace{-0.35cm} 
 &    \\
  \BraRat{\BdDecayNorm} &   $(\Err{1.27}{0.05}) \times 10^{-3}$   \\ 
 \BraRat{\BuDecayNorm} &  $(\Err{1.020}{0.019}) \times 10^{-3}$     \\
 \BraRat{\dbarkpicf}  &    $(\Err{3.947}{0.030}) \times 10^{-2}$   \\ 
  \BraRat{\dskkp} &   $(\Err{5.38}{0.10}) \times 10^{-2}$   \\
  \BraRat{\jpsimumu}  &    $(\Err{5.961}{0.033}) \times 10^{-2}$  \\
 $\BraRat{\Kstarz\to\Kp\pim}$ &   2/3   \\  
${f_{s}}/{f_{d}}$   &  $\Err{0.249}{0.020}$  \\
\end{tabular}
\label{tab:Branchingratios}
\end{table}

The acceptance, reconstruction and selection efficiencies are determined from simulation, with corrections applied for the particle identification response of the detector. The latter corrections are obtained as a function of muon and hadron track momentum and pseudorapidity, using a set of  calibration channels taken from data~\cite{LHCb-PUB-2016-021}. These give clean samples of particle species, independent of LHCb RICH- and muon-PID identification, from low-background decays such as $D^{*+}\to D^0\pi^+ , D^0\to K^-\pi^+$. 
For the branching fraction measurements of  \Bu modes and the \BcDecayJpsi decay, the efficiency of the normalisation mode is corrected for the reconstruction efficiency of the different number of tracks in the signal modes: three versus five, respectively. The correction and its systematic uncertainty are derived from tag-and-probe tracking efficiency measurements made as a function of track momentum and pseudorapidity~\cite{LHCb-DP-2013-002}.
The ratio of efficiencies is also corrected for differences in the trigger efficiency between simulation and data. These corrections are measured in data using a technique that compares the trigger efficiency when a signal candidate is used in the trigger decision with the case where the decision is taken independently of the signal candidate~\cite{LHCb-PUB-2014-039}.

All data at the three centre-of-mass energies are fitted together, and the final efficiency used in measuring the 
branching fraction for each mode is calculated from a luminosity-scaled combination of the efficiencies measured for each dataset, weighted by the ratio of the $b$-quark production cross-sections, given in Refs.~\cite{LHCb-PAPER-2016-031,LHCb-PAPER-2017-037}. 
After all contributions are considered, Table~\ref{tab:eventyields} shows the  fitted yields and efficiencies for the decays \GenDecay , \GenDecayJpsi  and the normalisation modes (before S-wave subtraction). The total efficiency is around 0.1\,\% for the signal decays. The efficiencies after selection of the normalisation modes are 0.6\,\% and 2.0\,\% for \BdDecayNorm and \BuDecayNorm decays, respectively. 
\begin{table}[h]
  \caption{
    \small 
   Fitted yields and efficiencies for the decays \GenDecay , \GenDecayJpsi  and the normalisation modes (before S-wave subtraction). The  quoted uncertainties arise solely from the limited size of the simulation samples.
    }
\begin{center}\begin{tabular}{lcc}
    Decay mode & Fitted yields  & Efficiency (\%)  \\ 
    \hline 
   \vspace{-0.35cm} 
  &  & \\
\BdDecay     & 0.3 $\pm$ 2.4  & 0.121  $\pm$ 0.003 \\
\BuDecay   & 0.0 $\pm$ 1.1  & 0.084 $\pm$ 0.002 \\
\BsDecay   & 0.0 $\pm$ 1.1  & 0.129 $\pm$ 0.004 \\
\BcDecay & 0.0 $\pm$ 1.1  & 0.025 $\pm$ 0.001 \\ 
    \hline
   \vspace{-0.35cm} 
  &  & \\
\BdDecayJpsi & 2.4 $\pm$ 2.9 & 0.115 $\pm$ 0.003 \\
\BuDecayJpsi & 0.2 $\pm$ 1.1 & 0.129   $\pm$ 0.006 \\
\BsDecayJpsi &  0.0 $\pm$ 1.1 & 0.098   $\pm$ 0.004 \\
\BcDecayJpsi & 127 $\pm$ 12 & 0.076 $\pm$ 0.002 \\
    \hline
   \vspace{-0.35cm} 
  &  & \\
\BdDecayNorm & $(9.53 \pm 0.02)\times 10^5$ & 0.649 $\pm$ 0.007 \\ 
\BuDecayNorm & $(3.56 \pm 0.03)\times 10^5$ & 1.979 $\pm$ 0.024 \\
  \end{tabular}\end{center}
\label{tab:eventyields}
\end{table}

\begin{table}[bt]
\centering
\caption{\small
Relative systematic uncertainties  for the branching-fraction measurements defined in \eqinline{eqn:BRCalc}, quoted separately for the different sources. The total is the sum in quadrature along the corresponding row.
}
\begin{tabular}{lccccc}
Measurement & External  & Simulation & PDF & Normalisation & Total \\
    &  [\%]  &  [\%]  &  [\%]  & procedure [\%]  & [\%]  \\
\hline 
\vspace{-0.35cm} 
   &   &   &   &   &   \\
\BraRat{\BdDecay}     & ~4.0 & 10.3 & 0.4 & - & 11.1 \\
\BraRat{\BdDecayJpsi} & ~4.0 & 11.8 & 0.3 & - & 12.5 \\
\BraRat{\BsDecay}     & ~8.8 & 11.1 & 0.3 & - & 14.2 \\
\BraRat{\BsDecayJpsi} & ~8.8 & 12.2 & 0.3 & - & 15.0 \\
\BraRat{\BuDecay}     & ~4.3 & \phantom{0}7.9 & 0.5 & 3.0 & \phantom{0}9.5 \\
\BraRat{\BuDecayJpsi} & ~4.3 & \phantom{0}9.9 & 0.7 & 3.0 & 11.2 \\
${f_{c}}/{f_{u}} \,\cdot\,$\BraRat{\BcDecay}     & 12.2 & \phantom{0}9.1 & 1.9 &  -  & 15.3 \\
${f_{c}}/{f_{u}} \,\cdot\,$\BraRat{\BcDecayJpsi} & ~4.3 & \phantom{0}5.9 & 1.2 & 3.0 & \phantom{0}8.0 \\
\end{tabular}
\label{tab:SystematicsSummary}
\end{table}

Many of the systematic uncertainties cancel in the ratio of efficiencies by construction of the signal branching fractions expressed in \eqinline{eqn:BRCalc}, nevertheless several systematic contributions remain, which are summarized in \tab{tab:SystematicsSummary}.  The first of these is the external uncertainty imported from the established branching fraction measurements~\cite{PDG2022}.
For the \Bs modes, the uncertainty of $f_s\big/f_d$ 
is also added in quadrature.  The \BcDecay signal channel has a larger   uncertainty due to the low signal yield of the normalisation-mode channel, however the yields of  the other normalisation samples incur negligible errors on the
branching-ratio measurements.
The second is the precision on the efficiencies used in \eqinline{eqn:BRCalc} which is a result of the uncertainties on the accuracy of the simulation's replication of the data. This systematic uncertainty is made up from several contributions added in quadrature. The finite size of the simulated sample limits the accuracy of the correction derived from them (2.1--4.7\,\% relative uncertainty). Only a subset of running conditions are simulated even though data from all data-taking conditions are used (2.7--5.0\,\% relative uncertainty). Possible mismodelling of the data leads to a 2.7--7.7\,\% relative systematic uncertainty. This uncertainty includes any imperfect BDT response estimation and also the efficiency variation over $q^2$ to account for different form-factor models and comparisons of angular distributions resulting from phase-space and spin-dependent decay of the di-muon system. The PID response in the simulation is corrected by resampling the calibration-channel data, and through the use of alternate sampling distributions, relative uncertainties of up to 3.1\,\% are estimated. The trigger efficiencies are also corrected for differences between simulation and data 
(3.2--7.0\,\% relative uncertainty). 
The third systematic contribution is a consequence of the low number of candidate events in the final fit requiring that most PDF parameters must be fixed from fits to simulation to ensure fit stability. Subsequently, these parameters are varied systematically by $\pm 1$ standard deviation and the fits repeated on data to assess the uncertainty from these shape parameters. The fits are also repeated with an alternative combinatorial background model (0.3--1.9\,\% total relative uncertainty).
The final systematic contribution is associated with the three modes where the corresponding normalisation channel has a different number of tracks from the signal. This includes the uncertainty in tracking efficiencies and differences in the interactions of charged pions and kaons with the detector material, added in quadrature (3.0\,\% combined relative uncertainty).

\section{Results}
\label{sec:Results}

As seen in Fig.~\ref{fig:BcMassFit}, a clear signal is observed in the  \BcDecayJpsi decay mode, and the corresponding branching fraction  is measured to be 
\begin{equation*}
	\frac{f_{c}}{f_{u}}\cdot
	\BraRat{\BcDecayJpsi} = \SFSplitErr{1.63}{0.15}{0.13}{-5}\,,
\end{equation*}
where the first uncertainty is statistical and the second is  systematic. This improves the precision of the first measurement of this decay, previously reported in Ref.~\cite{LHCb-PAPER-2013-010}.  As there is absence of signal in all other decay modes (less than 2 standard deviations),   the \cls method~\cite{CLs} is used to evaluate the compatibility of the observed invariant-mass distributions for each search mode with signal-and-background and background-only hypotheses. The distributions of {\it p}-values
as a function of assumed branching fraction are used to derive upper limits on the branching fractions at the 90\,\% and 95\,\%  CL, which are given in \tab{tab:UpperLimits}.

\begin{table}[htb]
  \caption{
    \small 
    Upper limits at the 90\,\% and 95\,\% confidence levels for \GenDecay and \GenDecayJpsi decays.
    }
\begin{center}\begin{tabular}{lcc}
    Branching fraction & \multicolumn{2} {c} {Upper limits} \\ 
     & 90\,\% CL & 95\,\% CL \\ 
    \hline 
   \vspace{-0.35cm} 
  &  & \\
    \BraRat{\BdDecay}     & \SF{4.0}{-8}  &  \SF{5.1}{-8} \\
    \BraRat{\BuDecay}   & \SF{2.4}{-8}  & \SF{3.2}{-8} \\
    \BraRat{\BsDecay}   & \SF{1.2}{-7} &  \SF{1.6}{-7} \\
    ${f_{c}}/{f_{u}} \,\cdot\,$\BraRat{\BcDecay} & \SF{7.5}{-8}  & \SF{9.6}{-8}\\ 
    \hline
   \vspace{-0.35cm} 
  &  & \\
    \BraRat{\BdDecayJpsi} & \SF{9.6}{-7} & \SF{1.1}{-6} \\
    \BraRat{\BuDecayJpsi} & \SF{2.8}{-7} & \SF{3.5}{-7} \\
    \BraRat{\BsDecayJpsi} & \SF{1.0}{-6} & \SF{1.5}{-6} 
  \end{tabular}\end{center}
\label{tab:UpperLimits}
\end{table}

Two measurements relating to \BcDecayPartReco decays are also extracted from the fit shown in \fig{fig:BcMassFit}.
The  ratio of yields \BcDecayPartReco to \BcDecayJpsi decays is measured to be  
\begin{equation*}
	\mathcal{R}_{D_s^{*+}/D_s^+}
 =\frac{\BraRat{\BcDecayPartReco}}{\BraRat{\BcDecayJpsi}} = \Err{1.91}{0.20} \pm {0.07},
\end{equation*}
where the systematic error is associated with the assumption that the efficiencies for the partially-reconstructed \BcDecayPartReco decay and the \BcDecayJpsi decay are equal.
The ratio of the number of \BcDecayPartReco decays described by the \Apm helicity amplitude compared to the total number is
\begin{equation*}
	\Gamma_{\pm\pm}/\Gamma_{\rm{tot}} = \frac{N_{\Apm}}{N_{\Apm} + N_{\Azz}} = 0.50\pm 0.11\pm 0.05 ,
\end{equation*}
where the uncertainty is dominantly statistical and the systematic term results from considering alternative fit models.    
The values of $\mathcal{R}_{D_s^{*+}/D_s^+}$  and $\Gamma_{\pm\pm}/\Gamma_{\rm{tot}}$   supersede  the previous \lhcb results \cite{LHCb-PAPER-2013-010},
and also are in agreement  with the ATLAS measurements \cite{ATLAS_2022_Bc2DsJpsi}. The ratio $\Gamma_{\pm\pm}/\Gamma_{\rm{tot}}$  is consistent with  the naive expectation of $2/3$ from spin-counting considerations.\\

\section{Conclusions}
\label{sec:Conclusions}

A search  for four rare \GenDecay decays 
is performed using proton-proton collision data collected by the \lhcb experiment  at centre-of-mass energies of 7, 8 and 13\tev, corresponding to a total integrated luminosity of 9\invfb. 
No new signals are observed and upper limits are set. 
Additional  limits are
determined when the muon pair originates from a \jpsimumu decay.
All  upper limits are either an improvement on existing results or are the first limits set by any experiment. Improved measurements of the previously observed \BcDecayJpsi decay are also made. 

In the future, it is expected that a new measurement of these modes with a 50\invfb dataset from the upgraded \lhcb experiment,  later increasing to 300\invfb with Upgrade~II,  will probe down towards the $10^{-9}$ level for the non-resonant modes, assuming luminosity scaling. These sensitivities will approach SM expectations, in particular for the \Bd modes studied in the present paper.

\section*{Acknowledgements}
%
%
\noindent We express our gratitude to our colleagues in the CERN
accelerator departments for the excellent performance of the LHC. We
thank the technical and administrative staff at the LHCb
institutes.
We acknowledge support from CERN and from the national agencies:
CAPES, CNPq, FAPERJ and FINEP (Brazil); 
MOST and NSFC (China); 
CNRS/IN2P3 (France); 
BMBF, DFG and MPG (Germany); 
INFN (Italy); 
NWO (Netherlands); 
MNiSW and NCN (Poland); 
MCID/IFA (Romania); 
MICINN (Spain); 
SNSF and SER (Switzerland); 
NASU (Ukraine); 
STFC (United Kingdom); 
DOE NP and NSF (USA).
We acknowledge the computing resources that are provided by CERN, IN2P3
(France), KIT and DESY (Germany), INFN (Italy), SURF (Netherlands),
PIC (Spain), GridPP (United Kingdom), 
CSCS (Switzerland), IFIN-HH (Romania), CBPF (Brazil),
Polish WLCG  (Poland) and NERSC (USA).
We are indebted to the communities behind the multiple open-source
software packages on which we depend.
Individual groups or members have received support from
ARC and ARDC (Australia);
Minciencias (Colombia);
AvH Foundation (Germany);
EPLANET, Marie Sk\l{}odowska-Curie Actions and ERC (European Union);
A*MIDEX, ANR, IPhU and Labex P2IO, and R\'{e}gion Auvergne-Rh\^{o}ne-Alpes (France);
Key Research Program of Frontier Sciences of CAS, CAS PIFI, CAS CCEPP, 
Fundamental Research Funds for the Central Universities, 
and Sci. \& Tech. Program of Guangzhou (China);
GVA, XuntaGal, GENCAT and Prog.~Atracci\'on Talento, CM (Spain);
SRC (Sweden);
the Leverhulme Trust, the Royal Society
 and UKRI (United Kingdom).



\addcontentsline{toc}{section}{References}
\bibliographystyle{LHCb}
\bibliography{main,standard,LHCb-PAPER,LHCb-DP}
 
\newpage
\centerline
{\large\bf LHCb collaboration}
\begin
{flushleft}
\small
R.~Aaij$^{32}$\lhcborcid{0000-0003-0533-1952},
A.S.W.~Abdelmotteleb$^{50}$\lhcborcid{0000-0001-7905-0542},
C.~Abellan~Beteta$^{44}$,
F.~Abudin{\'e}n$^{50}$\lhcborcid{0000-0002-6737-3528},
T.~Ackernley$^{54}$\lhcborcid{0000-0002-5951-3498},
B.~Adeva$^{40}$\lhcborcid{0000-0001-9756-3712},
M.~Adinolfi$^{48}$\lhcborcid{0000-0002-1326-1264},
P.~Adlarson$^{77}$\lhcborcid{0000-0001-6280-3851},
H.~Afsharnia$^{9}$,
C.~Agapopoulou$^{13}$\lhcborcid{0000-0002-2368-0147}
% ...

{\footnotesize \it

% ...
$^{9}$Universit{\'e} Clermont Auvergne, CNRS/IN2P3, LPC, Clermont-Ferrand, France\\
% ...
$^{13}$LPNHE, Sorbonne Universit{\'e}, Paris Diderot Sorbonne Paris Cit{\'e}, CNRS/IN2P3, Paris, France\\
% ...
$^{32}$Nikhef National Institute for Subatomic Physics, Amsterdam, Netherlands\\
% ...
$^{40}$Instituto Galego de F{\'\i}sica de Altas Enerx{\'\i}as (IGFAE), Universidade de Santiago de Compostela, Santiago de Compostela, Spain\\
% ...
$^{44}$Physik-Institut, Universit{\"a}t Z{\"u}rich, Z{\"u}rich, Switzerland\\
% ...
$^{48}$H.H. Wills Physics Laboratory, University of Bristol, Bristol, United Kingdom\\
% ...
$^{50}$Department of Physics, University of Warwick, Coventry, United Kingdom\\
% ...
$^{53}$School of Physics and Astronomy, University of Glasgow, Glasgow, United Kingdom\\
$^{54}$Oliver Lodge Laboratory, University of Liverpool, Liverpool, United Kingdom\\
% ...
$^{77}$Department of Physics and Astronomy, Uppsala University, Uppsala, Sweden, associated to $^{53}$\\
% ...
}
\end{flushleft}

\end{document}